\documentclass[preprint,showpacs,pre,preprintnumbers,amsmath,amssymb]{revtex4}

\newcommand{\commentoutA}[1]{}

\usepackage{graphicx}
\usepackage{dcolumn}
\usepackage{algorithm}
\usepackage{algorithmic}
\usepackage{bm}
\begin{document}

\preprint{LA-UR 15-20223}

\title{Canonical density matrix perturbation theory}

\author{Anders M.\ N. Niklasson \footnote{Corresponding author: amn@lanl.gov, anders.niklasson@gmail.com}} 
\affiliation{Theoretical Division, Los Alamos National Laboratory, Los Alamos, New Mexico 87545}
\author{Emanuel H. Rubensson, Elias Rudberg}
\affiliation{Division of Scientific Computing, Department of Information Technology, Uppsala University Box 337, SE-751 05 Uppsala, Sweden}
\author{M.\ J. Cawkwell}
\affiliation{Theoretical Division, Los Alamos National Laboratory, Los Alamos, New Mexico 87545}

\date{\today}

\begin{abstract}
Density matrix perturbation theory [Niklasson and Challacombe, Phys.\ Rev.\ Lett.\ {\bf 92}, 193001 (2004)]
is generalized to canonical (NVT) free energy ensembles in tight-binding, Hartree-Fock or Kohn-Sham density functional theory. 
The canonical density matrix perturbation theory can be used to calculate temperature dependent response properties 
from the coupled perturbed self-consistent field equations as in density functional 
perturbation theory.  The method is well suited to take advantage of sparse matrix algebra to achieve linear scaling
complexity in the computational cost as a function of system size for sufficiently large 
non-metallic materials and metals at high temperatures.
\end{abstract}

\maketitle

\section{Introduction}

Materials properties such as electric conductivity, magnetic susceptibility or electrical polarizabilities, 
are defined from their response to perturbations that are governed by the quantum nature of the electrons. 
The calculation of such quantum response properties represents a major challenge because of the high cost involved.
In traditional calculations the computational complexity scales cubically, ${\cal O}(N^3)$, or worse, with the number of atoms $N$, 
even when effective mean field models or density functional theory are used \cite{JPople79,SBaroni01}.  
By using the locality of the electronic solutions it is possible to reduce the computational cost 
for sufficiently large, non-metallic, materials to scale only linearly, ${\cal O}(N)$, 
with the system size  \cite{WYang91,GGalli96,POrdejon98,SGoedecker99,GScuseria99,SWu02,DBowler10,DBowler12}.
Initially, the development of linear scaling electronic structure theory was aimed at calculating ground state properties 
and not until recently has the focus shifted towards the computationally more demanding task of calculating the quantum response.  
A number of approaches to a quantum perturbation theory with reduced complexity have now been proposed and analyzed
\cite{SYokojima98,KYabana99,ATsolakidis02,MLazzeri03,COchsenfeld04,ANiklasson04,VWeber04,VWeber05,ANiklasson07C,AIzmaylov06,SCoriani07,FWang07}.  
Linear scaling quantum perturbation theory has so far mainly concerned properties at zero electronic temperature.
Here we extend the idea behind linear scaling density matrix perturbation theory \cite{ANiklasson04,VWeber04,VWeber05,ANiklasson07C} 
to calculations of static response properties valid also at finite electronic temperatures with fractional occupation of the states. 
Our proposed canonical density matrix perturbation theory, which is applicable within effective 
single-particle formulations, such as tight-binding, Hartree-Fock or Kohn-Sham density functional theory,
can be applied to calculate temperature dependent response properties from the solution of the coupled perturbed self-consistent 
field equations \cite{JPople79,HSekino86,SKarna91} as in density functional perturbation theory \cite{SBaroni87,SBaroni01}.
The canonical density matrix perturbation scheme should be directly applicable in 
a number of existing program packages for linear scaling electronic structure calculations, 
including CONQUEST \cite{EHernandez96,DBowler06,DBowler10}, CP2K \cite{Quickstep},
ERGO \cite{ERGO1,ERGO2}, FEMTECK \cite{ETsuchida98,ETsuchida07}, FreeON \cite{FreeON},
HONPAS \cite{HONPAS}, LATTE \cite{LATTE,MCawkwell12}, ONETEP \cite{NHine09}, OPEN-MX \cite{TOzaki05x}, and SIESTA \cite{JSoler02}.
While originally motivated by its ability to achieve linear scaling complexity, our canonical density matrix perturbation theory
is quite general and straightforward to use with high efficiency also for material systems that are too small to reach the
linear scaling regime. The computational kernel of the algorithm is centered around generalized matrix-matrix multiplications that
are well known to provide close to peak performance on many computer platforms using dense algebra,
including graphics processing units (GPU's) \cite{Cawkwell2012,Chow15}.

The paper is outlined as follows;
first we present the canonical density matrix perturbation theory. Thereafter we show how it can be used to
calculate temperature dependent free energy response properties, such as static polarizabilities and hyperpolarizabilities.
We discuss the alternative of using finite difference schemes and its potential problems. We conclude by discussing 
the capability of the canonical density matrix perturbation theory to reach linear scaling complexity
in the computational cost.

\section{Canonical density matrix perturbation theory}

In our density matrix perturbation theory we will use the single-particle density matrix 
and its derivatives to represent the electronic structure 
and its response to perturbations.  With the density matrix formulation it is easy to utilize matrix
sparsity from electronic nearsightedness \cite{WKohn96,SGoedecker99,MHastings04,VWeber05} and it allows direct calculations of observables.
The effective single-particle density matrix, $P$, at the electronic temperature $T_e$, can be calculated from the Hamiltonian, 
$H$, using a recursive Fermi operator expansion \cite{ANiklasson03B,ANiklasson08b,ANiklasson11,EHRubensson12},
\begin{equation}\label{Rec_F_1}\begin{array}{l}
{\displaystyle P = \left[e^{\beta(H-\mu I)} + I\right]^{-1} }\\
~~\\
{\displaystyle ~~~ \approx  {\cal F}_M({\cal F}_{M-1}(\ldots {\cal F}_0(H)\ldots ))},
\end{array}
\end{equation}
where the inverse temperature $\beta = 1/(k_B T_e)$, $\mu$ is the chemical potential, and $I$ is the identity matrix (see Appendix).
Both $H$ and $P$ are here assumed to be matrix representations in an orthogonal basis.
The expansion can be calculated through intermediate matrices $X_n = {\cal F}_n(X_{n-1})$ for $n = 1,2,3,\ldots, M$, where
\begin{equation}\begin{array}{l}\label{Rec_F_2}
{\displaystyle X_0 = {\cal F}_0(H) = \frac{1}{2}I - 2^{-(M+2)}\beta (H-\mu I),}\\
~~\\
{\displaystyle X_n = {\cal F}_{n}(X_{n-1}) = \frac{X_{n-1}^2}{X_{n-1}^2 + (I-X_{n-1})^2}}.
\end{array}
\end{equation}
In the canonical (NVT) ensemble, the chemical potential $\mu$ is chosen such that the density 
matrix has the correct occupation, $Tr[P] = N_{\rm occ}$, where $N_{\rm occ} $ is the number of occupied states.
The recursion scheme above provides a very efficient and rapidly converging expansion and 
the number of recursion steps $M$ can be kept low $(M < 20)$.  Because of the particular form of the 
Pad\'e polynomial ${\cal F}_n(X_{n-1})$, each iteration involves a solution of a system of linear equations,
which is well tailored for the linear conjugate gradient method \cite{ANiklasson03B,ANiklasson08b,EHRubensson12}. 
The recursive expansion avoids the calculation of individual eigenvalues and eigenfunctions and is therefore
well suited to reach linear scaling complexity in the computational cost for sufficiently 
large non-metallic problems, which can utilize thresholded sparse matrix algebra \cite{SGoedecker99}.

A canonical density matrix response expansion,
\begin{equation}\label{P_resp}
P(\lambda) = P^{(0)} + \lambda P^{(1)} + \lambda^2 P^{(2)} + \ldots~~~,
\end{equation}
where $Tr[P^{(k)}] = 0$ for $k>0$, 
with respect to a perturbation in the Hamiltonian, 
\begin{equation}\label{H_prt}
H(\lambda) = H^{(0)} + \lambda H^{(1)} + \lambda^2 H^{(2)} + \ldots,
\end{equation}
can be constructed at finite electronic temperatures, $T_e > 0$, based on the recursive 
Fermi operator expansion in Eqs.\ (\ref{Rec_F_1}) and (\ref{Rec_F_2}) above.
The technique is given by a free energy generalization 
of the zero temperature linear scaling density matrix perturbation theory \cite{ANiklasson04,VWeber04}.
The idea is to transfer the perturbations up to some specific order in each iteration step in the recursive 
Fermi-operator expansion, i.e.
\begin{equation}
X_{n}^{(k)} = \left. \frac{1}{k!}\frac{\partial^k}{\partial \lambda^k}{\cal F}_n(X_{n-1}^{(0)}+\lambda X_{n-1}^{(1)}+\ldots)\right\lvert_{\lambda = 0},
\end{equation}
for $n = 0,1,\ldots,M$, where $X_{-1}^{(k)} = H^{(k)}$.
The additional problem of conserving the number of particles in a canonical ensemble, which requires $Tr[P^{(k)}] = 0$ for $k>0$, 
is achieved by including the corresponding perturbative expansion of the chemical potential, i.e.
\begin{equation}
\mu = \mu(\lambda) = \mu^{(0)} + \lambda \mu^{(1)} + \lambda^2 \mu^{(2)} + \ldots ~.
\end{equation}
The values of $\mu^{(k)}$ ($k = 0,1,2,\ldots$) can be found by an iterative Newton-Raphson optimization 
of the occupation error with respect to the chemical potential using the relation
\begin{equation}\label{NewtRaphs}
{\displaystyle \left. \left(\frac{1}{\lambda^k}\frac{\partial P}{\partial \mu^{(k)}}\right)\right\lvert_{\lambda = 0} = P_\mu  = \beta P^{(0)}(I-P^{(0)})},
\end{equation}
which for the approximate expanded density matrix, Eqs.\ (\ref{Rec_F_1}) and (\ref{Rec_F_2}), is exact in the limit $M \rightarrow \infty$.
The trace of $P_\mu$, defined here, gives the change in occupation with respect to a change in $\mu$.
The small deviation from the exact analytic derivative for a finite expansion order $M$ is in practice insignificant, though for very low
values of $M$ the rate of convergence will be slightly lower than quadratic in analogy to quasi Newton schemes.
In combination with low temperatures, low values of $M$ may also lead to loss of convergence (see Table \ref{Tab_Conv}).
However, in this case we could typically use regular zero temperature response theory, or alternatively, 
a modified search routine to adjust for the correct occupation would be needed.

The canonical density matrix perturbation theory based on Eqs.\ (\ref{Rec_F_1}-\ref{NewtRaphs}) above, 
which is our first key result, is summarized by Algorithm\ \ref{Canonical_Fermi_Op_PRT} for up to
third order response.
Each inner loop requires the solution of a system of linear equations, which can be achieved with the conjugate gradient
method using $X_{n-1}^{(k)}$ as initial guesses. The linear conjugate gradient method \cite{GGolub96} is ideal for this purpose, since it 
efficiently can take advantage of matrix sparsity to reduce the scaling of the computational cost \cite{ANiklasson03B}. 
Generalizations and modifications to higher order response, grand canonical schemes (with a fixed value of $\mu$), or
spin-polarized (unrestricted) systems are straightforward.
It is interesting to note that the system matrices on the left hand-side of the inner loop of Algorithm\ \ref{Canonical_Fermi_Op_PRT}
are all the same, i.\ e. $T_{n-1}^{(0)}$. The same inverse of $T_{n-1}^{(0)}$ would therefore give the response $X^{(k)}_n$ for all 
orders $k$. The conditioning of the response algorithm should therefore be the same as for the original 0th-order expansion.
The system matrix $T_{n-1}^{(0)}$ is very well conditioned with a spectral condition number smaller than or equal to 2 \cite{EHRubensson12}
at any point of the algorithm.
In the limit of low temperature and high $n$, $T_{n-1}^{(0)} \rightarrow I$ and in the opposite limit of high temperatures
does the condition number go to 1 as $T_{n-1}^{(0)} \rightarrow I/2$. The well behaved conditioning is independent of 
the condition number of the Hamiltonian used in the initialization. 

\begin{algorithm}
\caption{Canonical density matrix response theory}
\label{Canonical_Fermi_Op_PRT}
\algsetup{indent=1em}
\begin{algorithmic}
\STATE $M \gets \mbox{Number of recursion steps}$
\STATE $\mu^{(0)} \gets \mbox{Initial guess}$
\STATE $\mu^{(i)} \gets 0  \mbox{ Initial guess} ~~ \{i = 1,2,3\}$
\STATE $\beta = 1/(k_BT_e) \gets \mbox{Choose temperature}$
\WHILE{Occupation error $>$ Tolerance}
 \STATE $X_0^{(0)} = \frac{1}{2}I - 2^{-(2+M)}\beta(H^{(0)}-\mu^{(0)} I)$
 \STATE $X_0^{(i)} = -2^{-(2+M)}\beta (H^{(i)} - \mu^{(i)}I), ~~ \{i = 1,2,3\}$
 \FOR{$n = 1:M$}
   \STATE {\bf solve for} $X_n^{(i)}, ~~~ \{i = 0,1,2,3\}$
   \STATE $T_{n-1}^{(0)} X_n^{(0)} = C_{n-1}^{(0)}$
   \STATE $T_{n-1}^{(0)} X_n^{(1)} = C_{n-1}^{(1)} + B_{n-1}^{(1)}X_n^{(0)}$
   \STATE $T_{n-1}^{(0)} X_n^{(2)} = C_{n-1}^{(2)} + B_{n-1}^{(2)}X_n^{(0)} + B_{n-1}^{(1)}X_n^{(1)}$
   \STATE $T_{n-1}^{(0)} X_n^{(3)} = C_{n-1}^{(3)} + B_{n-1}^{(3)}X_n^{(0)} + B_{n-1}^{(2)}X_n^{(1)} + B_{n-1}^{(1)}X_n^{(2)}$
 \ENDFOR
 \STATE $P^{(i)} = X_M^{(i)}, ~~ \{i = 0,1,2,3\}$
 \STATE $\mu^{(0)} = \mu^{(0)} + (N_e - Tr[P^{(0)}])/Tr[P_\mu]$
 \STATE $\mu^{(i)} = \mu^{(i)} - Tr[P^{(i)}]/Tr[P_\mu], ~~ \{i = 1,2,3\}$
 \STATE ${\rm Occupation ~ error} = |Tr[P^{(0)}] - N_e| +  \sum_{i = 1}^3 |Tr[P^{(i)}]|$
\ENDWHILE
\STATE {\bf using:}
\STATE $P_\mu = \beta P^{(0)}(I-P^{(0)})$
\STATE $T_{n}^{(0)} = 2X_{n}^{(0)}(X_{n}^{(0)}-I) + I$
\STATE $C_{n}^{(m)} = \sum_{i+j = m} X_{n}^{(i)}X_{n}^{(j)}, ~~ \{i,j \ge 0, ~~ m = 0,1,2,3\}$
\STATE $B_{n}^{(m)} = 2(X_{n}^{(m)}-C_{n}^{(m)}), ~~\{m = 0,1,2,3\}$
\end{algorithmic}
\end{algorithm}

\section{Free energy response theory} 

To study the quantum response valid at finite electronic temperatures, the electronic
entropy contribution to the free energy has to be considered. We will look
at two different situations: a) non self-consistent band energy response
as in regular tight-binding theory using an orthogonal matrix representation
and b) self-consistent free energy response
as in density functional or Hartree-Fock theory using a non-orthogonal formulation. 
To clearly separate the two cases we will use two different notations. 
For the orthogonal tight-binding like formulation we keep using 
$H$ and $P$, which is consistent with the previous sections, and for 
the self-consistent free energy response we use $F$ and $D$ for
the non-orthogonal matrix representations and $F^\perp$ and $D^\perp$
for the orthogonalized representations, as is explained in the sections below.

\subsection{Non self-consistent tight-binding-like free energy response}

In a simple tight-binding like formulation, the expansion terms for the canonical free energy,
\begin{equation}\label{FreeE} \begin{array}{l}
\Omega(\lambda) = Tr[P(\lambda)H(\lambda)] - T_e{\cal S}[P(\lambda)] = \\
~\\
~~ = \Omega^{(0)} + \lambda \Omega^{(1)} + \lambda^2 \Omega^{(2)} + \ldots,
\end{array}
\end{equation}
generated by a perturbation in $H(\lambda)$, Eq.\ (\ref{H_prt}),
with the electronic entropy \cite{RParr89,ANiklasson08b},
\begin{equation}\label{SED}
{\cal S}[P] = - k_B Tr[P\ln(P) + (I-P)\ln(I-P)],
\end{equation}
are given by
\begin{equation}\label{Resp_0}
\Omega^{(m)} = \frac{1}{m} \sum_{k = 1}^{m} k Tr[H^{(k)}P^{(m-k)}].
\end{equation}
This expression, with $P^{(k)}$ calculated from our canonical density matrix perturbation scheme in
Algorithm \ref{Canonical_Fermi_Op_PRT}, is a straightforward generalization of the conventional $T_e = 0$ limit
of the ``$n+1$'' rule \cite{ANiklasson07C} and follows directly from the fact that the first order
response term $Tr[H^{(0)}P^{(1)}]$ is cancelled by the response in the entropy \cite{ANiklasson08b}. 
Higher-order derivatives of order $n+1$ therefore contain at most a derivative of order $n$ in the density matrix.
This generalization is possible only by including
the entropy term in Eq.\ (\ref{FreeE}), which is required to provide a variationally 
correct description of the energetics. We have not been able to find any explicit density matrix expressions for 
Wigner's $2n+1$ rule \cite{EAHylleraas30,EWigner35,RMcWeeny62,THelgaker02,ANiklasson04,VWeber05,JKussmann07} that are valid 
also at finite temperatures. A more detailed derivation of Eq (\ref{Resp_0}) is given in the appendix.

\subsection{Self-consistent free energy response}

In self-consistent first principles approaches such as Hartree-Fock theory \cite{RMcWeeny60} 
(density functional and self-consistent tight-binding theory, although different, 
follow equivalently) the free energy in the restricted case (without spin polarization) 
is given by a constrained minimization of the functional
\begin{equation}\label{FreeE_tot}
\Omega_{\rm SCF}[D] = 2Tr[hD] + Tr[DG(D)] - 2 T_e {\cal S}[D^\perp ],
\end{equation}
under the condition that $2Tr[DS] = N_e$, where $N_e$ is the number of electrons 
(two in each occupied state).  
Here $D^\perp$ is the orthogonalized representation of the Hartree-Fock density matrix $D$ 
such that $D = ZD^\perp Z^T$, and the orthogonalized effective single-particle Hamiltonian is given by $F^\perp = Z^TFZ$,
where the Fockian $F = h + G(D)$ and $Z$ is the inverse factor of the basis set overlap matrix $S$ such that $Z^TSZ = I$.
The density matrix, $D$, is thus given by
\begin{equation}\label{Fermi_D}
D = Z \left[e^{\beta(F^\perp-\mu I)} + I\right]^{-1} Z^T,
\end{equation}
which can be calculated through the recursive Fermi operator expansion in Eqs.\ (\ref{Rec_F_1}) and (\ref{Rec_F_2}). Here
$h$ is the usual one-electron term and $G(D)$ is the conventional two-electron part including
the Coulomb $J$ and exchange term $K$, respectively \cite{RMcWeeny60}. In density functional theory, the Fockian $F$ is replaced by
the corresponding Kohn-Sham Hamiltonian, where the exchange term $K$ is substituted with the exchange-correlation
potential term. Notice that to make a clear distinction to the non-self-consistent response we use 
the notation $D$ and $F$ for the self-consistent Hartree-Fock density matrix and Fockian, i.e. the 
effective single-particle Hamiltonian.

With a basis-set independent first order perturbation in the one-electron term,
\begin{equation}\label{FirstOrderPRT}
h(\lambda) = h^{(0)} + \lambda h^{(1)},
\end{equation}
for example due to an external electric field, the self-consistent response in the density matrix 
is given by the solution of the coupled perturbed self-consistent field (SCF) equations 
as in density functional perturbation theory:
\begin{equation}\begin{array}{l}
{\displaystyle F(\lambda) = h^{(0)} + \lambda h^{(1)} + G(D^{(0)} + \lambda D^{(1)} + \ldots)},\\
~~\\
{\displaystyle F^{\perp}(\lambda) = Z^T F(\lambda)Z},\\
~~\\
{\displaystyle D(\lambda) = Z\left[e^{\beta(F^{\perp}(\lambda)-\mu I)} + I\right]^{-1}Z^T},\\
\end{array}
\end{equation}
where $D$ and $F$ are expanded in terms of $\lambda$, i.e.
\begin{equation}\begin{array}{l}
{\displaystyle D(\lambda) = D^{(0)} + \lambda D^{(1)} + \lambda^2 D^{(2)} + \ldots ~,}\\
~\\
{\displaystyle F(\lambda) = F^{(0)} + \lambda F^{(1)} + \lambda^2 F^{(2)} + \ldots ~}.
\end{array}
\end{equation}
The coupled response equations above are solved in each iteration
using the canonical density matrix perturbation theory as implemented in Algorithm \ref{Canonical_Fermi_Op_PRT}
with $H$ and $P$ replaced by $F^\perp$ and $D^\perp$.
At self consistency, the free energy expansion terms,
\begin{equation}
\Omega_{\rm SCF}(\lambda) = \Omega_{\rm SCF}[D^{(0)}] + \lambda \Omega_{\rm SCF}^{(1)} + \lambda^2 \Omega_{\rm SCF}^{(2)} +  \ldots ~~,
\end{equation}
are given by
\begin{equation} \label{Resp_Omega}
\Omega_{\rm SCF}^{(m)} = \frac{2}{m} Tr[h^{(1)}D^{(m-1)}] ~~~ m > 0.
\end{equation}
This simple and convenient expression for the basis-set independent free energy response, 
which follows (see Appendix) from Eq.\ (\ref{Resp_0}), is another key result of this paper.
The free energy response theory presented here provides a general technique 
to perform reduced complexity calculations of, for example, temperature-dependent static 
polarizabilities and hyperpolarizabilities \cite{VWeber04,VWeber05}. 

\section{Finite difference approximations}

\begin{figure}[t]
\resizebox*{4.1in}{!}{\includegraphics[angle=00]{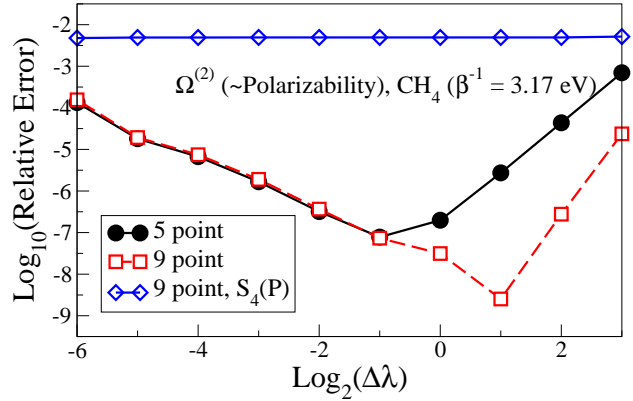}}\vspace{-1.5cm} 
\caption{\label{Fig_1}
The relative error compared to the ``exact'' derivative in Eq.\ (\ref{Resp_Omega}) for $\Omega_{\rm SCF}^{(2)}$ using
5 and 9 point central difference schemes for the calculation of the second order response in the free energy with 
respect to an external electric field, i.e.\ the polarizability. 
Either the exact entropy expression was used, Eq.\ (\ref{SED}), or the highest order ($m = 4$) approximation in Eq.\ (\ref{S_Approx}).
The electronic temperature $T_e$ is about 37,000 K.}
\end{figure}

An alternative to the canonical density matrix perturbation theory is to perform calculations 
with finite perturbations and use finite difference approximations of the free energy derivatives. 
However, this can be far from trivial because the numerical 
errors are sometimes difficult to estimate and control, in particular for high temperature 
hyperpolarizabilities.
Nevertheless, by using finite steps $\delta \lambda$ of the perturbations in $h$, combined 
with multi-point high-order finite difference schemes, it is sometimes possible to reach
good accuracy. This is illustrated in Fig.\ \ref{Fig_1}, which shows the finite difference
error in the approximation of the second order free energy response, $\Omega_{\rm SCF}^{(2)}$, 
with respect to an external electric field for a self-consistent tight-binding model 
\cite{MElstner98,MFinnis98,TFrauenheim00,BAradi07}
as implemented in the electronic structure program package LATTE \cite{LATTE,MCawkwell12}. 
Finite difference calculations of higher-order hyperpolarizabilities show similar behavior.

In a finite difference approximation it is difficult to know {\em a priori} 
what step size $\Delta \lambda$ to use for the perturbations 
$\lambda h^{(1)}$ in Eq.\ (\ref{FirstOrderPRT}). 
Errors may be large unless careful numerical testing is performed. 
This can be expensive and even when an optimal step size has 
been found, the computational cost is still higher than the analytical approach. For example, to calculate the second order
response using the five point finite difference scheme has a computational cost of about 5 times
a ground state calculation, whereas the cost for the density matrix perturbation theory is 
only about 3 times larger.  This cost estimate does not include the additional entropy calculations.
The calculation of the entropy is difficult (or impossible) to perform accurately within 
linear scaling complexity. Computationally favorable
formulations that are based on approximate expansions of ${\cal S}[P]$ in Eq.\ (\ref{SED}) are typically poor. 
For example, when any of the approximate entropy expressions,
\begin{equation}\label{S_Approx}
{\cal S}_m[P] \approx -k_B \sum_{i = 1}^{m} c_i(m) Tr[P^m(P-I)^m],
\end{equation}
with the coefficients $c_i(m)$ in Tab.\ \ref{Tab_Coef} are used,
the relative error of the polarizability in Fig.\ \ref{Fig_1}
is increased by over 6 orders of magnitude for the most accurate 9 point finite difference approximation.
The accuracy is at best only about 0.5 percent with any of the entropy approximations in Eq.\ (\ref{S_Approx})
and Tab.\ \ref{Tab_Coef}.  Only by avoiding explicit entropy calculations it is possible
to reach a meaningful accuracy. This is possible in a finite difference approximation by
using the finite differences of the dipole moments instead of the free energies.
Such calculations (not shown) avoid calculating the explicit entropy term and the numerical accuracy is 
similar to the finite difference approximations using the free energies with the exact entropy expression
as illustrated in Fig.\ \ref{Fig_1}.

\begin{table}[t]
  \centering
  \caption{\protect Coefficients for the approximate entropy expression in Eq.\ (\ref{S_Approx}). The coefficients are
  determined from the ansatz in Eq.\ (\ref{S_Approx}) with the requirement that the function value and a few of its
  derivatives are correct at the midpoint 0.5 of the interval [0,1] in which $P$ has its eigenvalues.
  }\label{Tab_Coef}
  \begin{ruledtabular}
  \begin{tabular}{clll}
    $c_i(m)$ & $m = 1$ & $m=2$ & $m = 4$ \\
     \hline
     $c_1(m)$  & $4\ln(2)$ &  $8\ln(2)-2$    & $16\ln(2)-34/5$    \\
     $c_2(m)$  &           &  $16\ln(2)-8$   & $96\ln(2)-844/15$   \\
     $c_3(m)$  &           &                 & $256\ln(2)-2336/15$  \\
     $c_4(m)$  &           &                 & $256\ln(2)-2368/15$     \\
  \end{tabular}
  \end{ruledtabular}
\end{table}

\section{first principles results}

\subsection{Polarizabilities and hyperpolarizabilities}

\begin{figure}[t]
\resizebox*{3.5in}{!}{\includegraphics[angle=00]{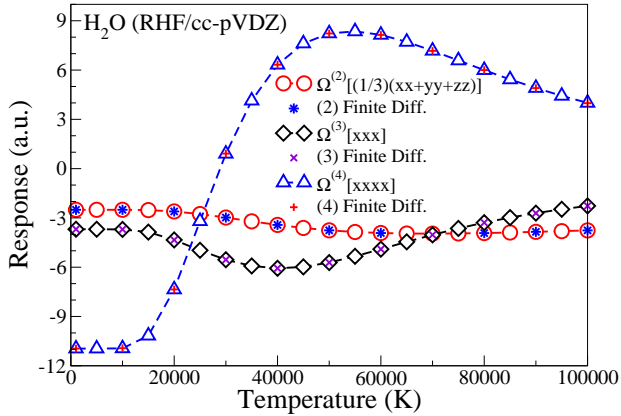}}\vspace{0.0cm}
\caption{\label{Fig_Water}
(color online) The temperature dependent isotropic second-order response
$\Omega_{\rm SCF}^{(2)}[\frac{1}{3}(xx+yy+zz)] = \frac{1}{3}(\Omega_{\rm SCF}^{(2)}[xx]
+\Omega_{\rm SCF}^{(2)}[yy] +\Omega_{\rm SCF}^{(2)}[zz])$, and the third order and fourth order
response in the $x$ direction. The canonical density matrix response Algorithm \ref{Canonical_Fermi_Op_PRT} 
for restricted Hartree-Fock theory (RHF) with a Gaussian basis set (cc-pVDZ) was used.
The xyz-coordinates of the molecule: 
$\{ {\rm O}~ (0.0, 0.0, 0.0); ~~{\rm H}~ (-1.809, 0.0, 0.0); ~~{\rm H} ~ (0.453549, 1.751221, 0.0) \}$ in atomic units.
As a comparison and validation we show five-point finite difference
calculations of the free energy derivatives.
At low electronic temperatures the second-order response corresponds to $1/2$ times 
the isotropic polarizability (see Tab.\ \ref{Tab_Conv}).}
\end{figure}

Figure \ref{Fig_Water} shows the calculated temperature-dependent response
for a single water molecule with respect to static electric fields. 
The calculations were performed with Hartree-Fock theory using the ERGO program package \cite{ERGO1,ERGO2}.
At lower temperatures the response 
values correspond to the isotropic polarizability and hyperpolarizabilities if the values are multiplied by $m!$,
i.\ e.\ the factorial of the response order.
At higher temperatures this interpretation is less accurate because of the limited
basis set description of the thermally excited states. For relevant temperatures
below 10,000 K our calculations show a very small temperature dependence,
which is consistent with a fairly large HOMO-LUMO gap.
For higher temperatures the errors may be significant, since the Gaussian basis set 
used here (cc-pVDZ) was not designed for high-temperature expansions.
The calculations were performed for a single molecule in the gas phase.
For periodic boundary conditions the position and the dipole moment 
operator are not well defined. In this case the techniques developed within
the modern theory of polarizability can be applied \cite{RDKingSmith93,RResta94,RResta98,Xiang06}.

The response properties converges quickly as a function of the number of recursion steps ($M$) in
the canonical density matrix response expansion in Alg. \ref{Canonical_Fermi_Op_PRT}, which is
illustrated in Tab.\ \ref{Tab_Conv}. At higher temperatures we see a slightly slower convergence,
and at low temperatures and with a small number of recursion steps there can be problems with convergence
of the occupation, since the chemical potential derivative estimate $P_\mu = \beta P^{(0)}(I-P^{(0)})$ is less
accurate. In this case we may prefer to use a regular zero-temperature response calculation.

\begin{table}[t]
  \centering
  \caption{\protect Convergence of the isotropic polarizability 
  $\alpha_{\rm iso} = 2\frac{1}{3}(\Omega^{(2)}_{\rm SCF}[xx] + \Omega^{(2)}_{\rm SCF}[yy] + \Omega^{(2)}_{\rm SCF}[yy])$ 
  for three different electronic temperatures $T_e$ (1000 K, 30,000 K and 100,000 K) as a function of the number of recursion steps ($M$) in 
  the canonical density matrix response expansion in Algorithm \ref{Canonical_Fermi_Op_PRT} for a 
  water molecule calculated from restricted Hartree-Fock theory (RHF) with a Gaussian basis set (cc-pVDZ).
  The xyz-coordinates of the molecule: 
  $\{ {\rm O}~ (0.0, 0.0, 0.0); ~~{\rm H}~ (-1.809, 0.0, 0.0); ~~{\rm H} ~ (0.453549, 1.751221, 0.0) \}$ in atomic units.
  As a comparison and validation we show the $T_e = 0$ K result of the isotropic polarizability, which was calculated 
  by solving the linear response time-dependent Hartree-Fock (or RPA) equations \cite{JOlsen88} as implemeted
  in the ERGO program package \cite{ERGO1,ERGO2} applied for the zero-frequency case.}
  \label{Tab_Conv}
  \begin{ruledtabular}
  \begin{tabular}{lcllcllcl}
     $T_e$ (K) & $M$ & $\alpha_{\rm iso}$ (a.u.) & $T_e$ (K) & $M$ & $\alpha_{\rm iso}$ (a.u.) & $T_e$ (K) & $M$ & $\alpha_{\rm iso}$ (a.u.) \\
     \hline
     0 (ERGO)  & n/a    &  -5.0112528623  &        &    &               &         &    &               \\
     1000  &     6      &  no convergence & 40,000 & 6  & -6.8540449154 & 100,000 & 6  & -7.5204026148 \\
     1000  &     8      &  -5.0112527697  & 40,000 & 8  & -6.8538983381 & 100,000 & 8  & -7.5198385798 \\
     1000  &     10     &  -5.0112527697  & 40,000 & 10 & -6.8538891617 & 100,000 & 10 & -7.5198033131 \\
     1000  &     12     &  -5.0112527697  & 40,000 & 12 & -6.8538885881 & 100,000 & 12 & -7.5198011089 \\
     1000  &     14     &  -5.0112527697  & 40,000 & 14 & -6.8538885522 & 100,000 & 14 & -7.5198009711 \\
     1000  &     16     &  -5.0112527697  & 40,000 & 16 & -6.8538885500 & 100,000 & 16 & -7.5198009625 \\
  \end{tabular}
  \end{ruledtabular}
\end{table}

\begin{figure}[t]
\resizebox*{3.7in}{!}{\includegraphics[angle=00]{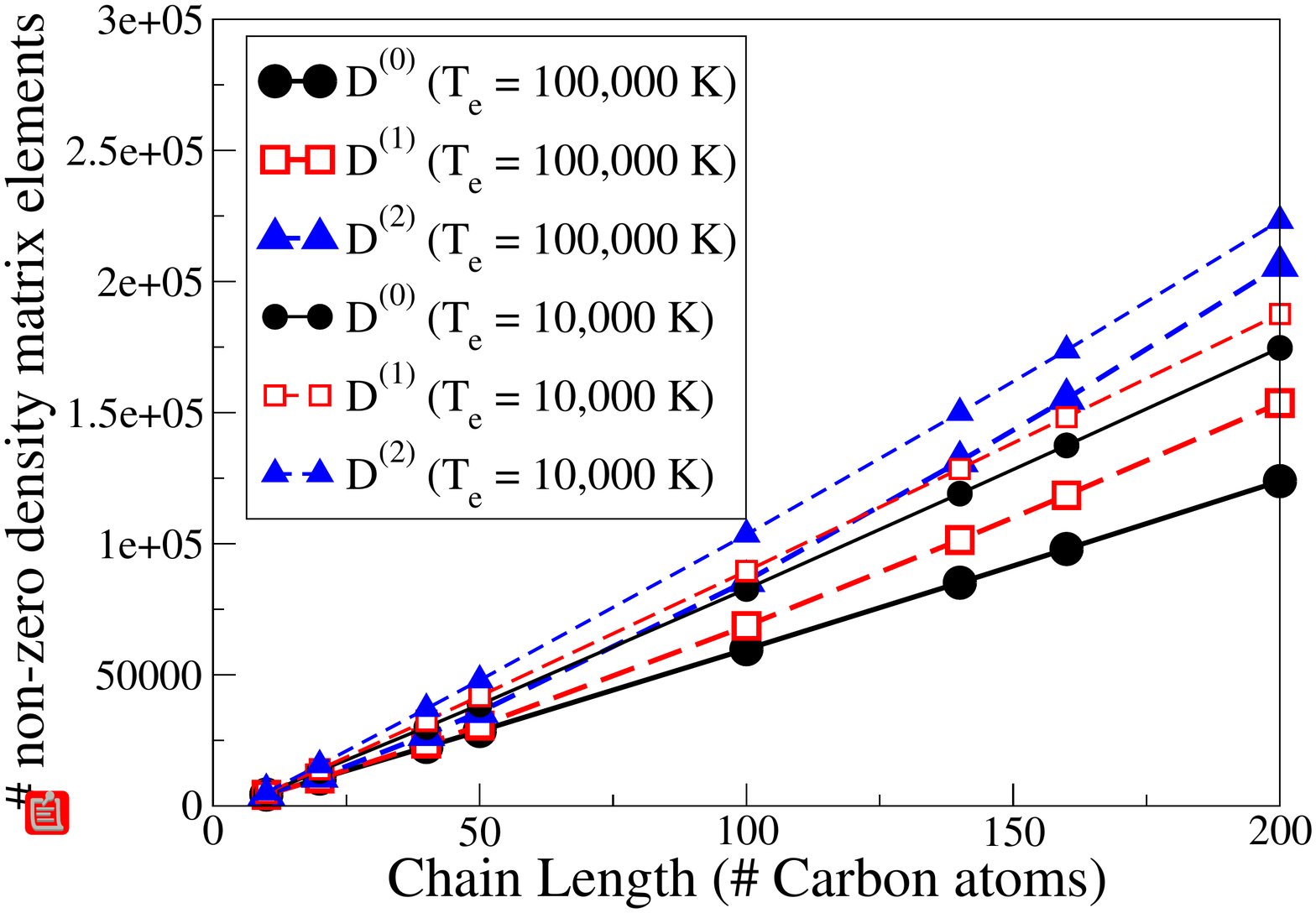}}\vspace{0.0cm}
\caption{\label{Fig_Scaling}
(color online) The sparsity scaling as a function of system size of the density matrix and its 
first and second order response with respect to an electric (static) dipole perturbation for two different electronic temperatures. 
The graphs show the number of non-zero elements of the orthogonal density matrix after a numerical threshold of 10$^{-5}$.}
\end{figure}

\subsection{Linear scaling complexity}

It is easy to understand the potential for a linear scaling implementation of canonical density matrix perturbation theory.
Owing to nearsightedness \cite{WKohn96,MHastings04,VWeber05,SGoedecker99}, both the Hamiltonian and its perturbations, as well as
the density matrix and its response, have sparse matrix representations for 
non-metallic materials when local basis set representations are used.
The number of significant matrix elements above some small numerical threshold (or machine precision) then scales only linearly 
with the number of atoms for sufficiently large systems. In this case, since all operations
in the canonical density matrix perturbation scheme in Algorithm \ref{Canonical_Fermi_Op_PRT} are based on matrix-matrix operations,
the computational cost scales only linearly with system size if sparse matrix
algebra is used in the calculations. This is not possible in regular Rayleigh-Schr\"{o}dinger
perturbation theory, which requires the calculation of individual
eigenvalues and eigenfunctions. Figure \ref{Fig_Scaling} shows the number of non-zero elements
above threshold as a function of system size for the density matrix and its first and second order response
with respect to an electric dipole perturbation.
The test systems are simple one-dimensional hydrocarbon chains of various lengths and
the calculations where performed based on Hartree-Fock theory using a small Gaussian (STO-3G) basis.
Gaussian basis sets were not designed for the high-temperature expansions demonstrated here
and we can expect that the accuracy is limited. The simulations therefore only serve as
a schematic demonstration of the expected behavior.
For example, at higher temperatures the locality, i.e.\ the matrix sparsity, is increased similar 
to what is found for larger HOMO-LUMO gaps \cite{SGoedecker99}, 
and for higher order response the locality decreases, as has been seen in previous studies of the
zero temperature case \cite{VWeber04,VWeber05}.
Using a larger Gaussian basis set should not change this general behavior of the locality
and the results would still be uncertain.
Matrix sparsity may also suffer and numerical problems may arise due to ill-conditioning
from linear dependencies between many Gaussians. However, this will not affect the conditioning
of the canonical density matrix response algorithm, Alg.\ \ref{Canonical_Fermi_Op_PRT}, and 
the low spectral condition number of $T^{(0)}_{n-1}$, which is always $<2$, 
but it would affect the congruence transformation from the non-orthogonal atomic orbital
representation of $F$ to $F^\perp$. The input data of the response algorithm would thus
be less accurate. Localized numerical atomic orbital basis sets that have been tailored specifically 
for high-temperature expansions (and with low condition numbers of the overlap matrix) would 
then be a more appropriate choice.

\section{Summary}

In summary, we have presented a canonical single-particle density matrix perturbation scheme that enables the
calculation of temperature dependent quantum response properties. Since our approach avoids the calculation
of individual eigenvalues and eigenfunctions as well as the entropy, the theory is well adapted for 
reduced complexity calculations with a computational effort that scales only linearly with the system size. 
However, we may expect very fast parallel performance also for smaller systems in the limit of dense matrix algebra,
since the computational kernel is centered around matrix-matrix multiplications that often can reach close
to peak performance on modern hardware.
The perturbation scheme should be applicable to a number of existing program packages for linear
scaling electronic structure calculations.  

\section{Acknowledgement}

The Los Alamos National Laboratory is operated by Los
Alamos National Security, LLC for the NNSA of the US-DoE
under Contract No. DE-AC52-06NA25396. We gratefully
acknowledge the support of the United States Department of
Energy (U.S. DOE) through LANL LDRD program, 
the G\"{o}ran Gustafsson Foundation, 
the Swedish research council (grant no. 621-2012-3861), and 
the Swedish national strategic e-science research program (eSSENCE) as well 
as stimulating contributions from Travis Peery 
at the T-Division Ten-Bar Java Group.

\section{Appendix}

\subsection{Recursive Fermi operator expansion}

There are several techniques to calculate matrix exponentials. For example, if we start with
\begin{equation}
e^x = \left(e^{x/n}\right)^n =  \left( \frac{e^{x/(2n)}}{e^{-x/(2n)}}\right)^n,
\end{equation}
a first order Taylor expansion gives
\begin{equation}
e^x = \lim_{n \rightarrow \infty} \left( \frac{2n+x}{2n-x}\right)^n.
\end{equation}
Using this expansion we can approximate the Fermi-Dirac distribution function, 
${\Phi}(x)$, with
\begin{equation}
{\Phi}(x) = [e^x + 1]^{-1} = \lim_{n \rightarrow \infty} \frac{(2n-x)^n}{(2n+x)^n + (2n-x)^n},
\end{equation}
such that
\begin{equation}
{\Phi}(2n - 4nx) \approx \frac{x^n}{x^n + (1-x)^n},
\end{equation}
which is accurate for large values of $n$.  The Pad\'e polynomial function
\begin{equation}
{f}_n(x) = \frac{x^n}{x^n + (1-x)^n}
\end{equation}
can be expanded recursively, since
\begin{equation}
{f}_{m \times n} (x) = {f}_m({f}_n(x)).
\end{equation}
This particular property enables a rapid high-order expansions in only a few iterations
in the recursive Fermi-operator expansion, 
\begin{equation}\begin{array}{l}
{\Phi}\left[\beta(\varepsilon_i-\mu)\right] = {\Phi}(2n - 4nx_i) \\
~~\\
\approx {f}_{n}(x_i) = {f}_2({f}_2( \ldots {f}_2(x_i) \ldots )),
\end{array}
\end{equation}
where
\begin{equation}
x_i = \frac{1}{2} - \frac{\beta}{4n}(\varepsilon_i -\mu)
\end{equation}
with the recursion repeated $m$ times, i.\ e.\ for $n = 2^m$.
In 30 steps ($m=30$) this gives an expansion order of the Pad\'e polynomial of over 1 billion,
but often less than 10 steps are needed.

The density matrix at finite electronic temperatures,
\begin{equation}
P = \left[e^{\beta(H-\mu I)} + 1\right]^{-1} = {\Phi}\left[\beta(H-\mu I)\right],
\end{equation}
can now be calculated with the recursive grand canonical Fermi operator expansion,
\begin{equation}
P = {f}_2\left(\ldots f_2\left({f}_2\left(\frac{1}{2}I - 2^{-(2+m)}\beta(H-\mu I)\right)\right) \ldots\right),
\end{equation}
which forms the starting point in Eq.\ (\ref{Rec_F_2}), with 
${\cal F}_n(X) = f_2(X)$. The recursive grand canonical Fermi operator expansion,
derivations, convergence analysis, and tests with various basis sets have been published previously in 
Refs.\ \cite{ANiklasson03B,ANiklasson08b,ANiklasson11,EHRubensson12}.

\subsection{Perturbation response for the non-self-consistent single particle free energy}

To derive Eq.\ (\ref{Resp_0}) we start by noting that from the definition of the
density matrix response and the perturbations in the Hamiltonian, Eqs.\ (\ref{P_resp}) and (\ref{H_prt}), 
we have 
\begin{equation}
\left. \frac{\partial^k}{\partial \lambda^k} P(\lambda)\right\lvert_{\lambda = 0} = P^{[k]} = k!P^{(k)}
\end{equation}
and
\begin{equation}
\left. \frac{\partial^k}{\partial \lambda^k} H(\lambda)\right\lvert_{\lambda = 0} = H^{[k]} = k!H^{(k)},
\end{equation}
where we use square brackets for the regular Taylor expansion terms, $H^{[k]}$ and $P^{[k]}$, and
round brackets, $H^{(k)}$ and $P^{(k)}$, for the perturbation expansions as in Eqs.\ (\ref{P_resp}) and (\ref{H_prt}).
Thereafter we can calculate the response terms $\Omega^{(m)}$ from the derivatives of
the free energy expression in Eq.\ (\ref{FreeE}), i.\ e.
\begin{equation}
\Omega^{(m)} = \left. \frac{1}{m!} \frac{\partial^m}{\partial \lambda^m} \Omega(\lambda)\right\lvert_{\lambda = 0}.
\end{equation}
It is easy to see that the first derivative of the entropy term ${\cal S}[P(\lambda)]$ in Eq.\ (\ref{SED}) 
is given by
\begin{equation}\label{EntDer}\begin{array}{l}
\left. \frac{\partial }{\partial \lambda} {\cal S}[P(\lambda)]\right\lvert_{\lambda = 0} = - k_B Tr\left[(\ln(P) -\ln(I-P))P^{[1]}\right] \\
= - k_B Tr\left[\ln\left(P(I-P)^{-1}\right)P^{[1]}\right] = -k_B Tr\left[ \ln(e^{-\beta(H-\mu I)})P^{[1]}\right] \\
= k_B \beta Tr[HP^{(1)}],
\end{array}
\end{equation}
since we have a canonical perturbation $Tr[P^{[1]}] = 0$ and $P = [e^{\beta(H-\mu I)}+I]^{-1}$.
This means that the first order response in the free energy $\Omega(\lambda)$ is given by
\begin{equation}\begin{array}{l}
\Omega^{(1)} = \left. \frac{\partial}{\partial \lambda} \Omega(\lambda)\right\lvert_{\lambda = 0} 
= Tr[H^{(1)}P^{(0)}] + Tr[H^{(0)}P^{(1)}] - T_e k_B \beta Tr[HP^{(1)}] \\
= Tr[H^{[1]}P^{[0]}] = Tr[H^{(1)}P^{(0)}].
\end{array}
\end{equation}
For the second order expansion we find that 
\begin{equation} \begin{array}{l}
\Omega^{(2)} = \frac{1}{2} Tr[H^{[1]}P^{[1]} + H^{[2]}P^{[0]}] \\
= \frac{1}{2} \left(Tr[H^{(1)}P^{(1)}] + 2Tr[H^{(2)}P^{(0)}]\right).
\end{array}
\end{equation}
For the third order expansion we find that
\begin{equation} \begin{array}{l}
\Omega^{(3)} = \frac{1}{6} Tr[H^{[1]}P^{[2]} + H^{[2]}P^{[1]} + H^{[3]}P^{[0]} + H^{[2]}P^{[1]}] \\
= \frac{1}{6} Tr[2H^{(1)}P^{(2)} + 2H^{(2)}P^{(1)} + 6H^{(3)}P^{(0)} + 2H^{(2)}P^{(1)}] \\
= \frac{1}{3} \left(Tr[H^{(1)}P^{(2)}] + 2Tr[H^{(2)}P^{(1)}] + 3Tr[H^{(3)}P^{(0)}]\right).
\end{array}
\end{equation}
The straightforward $m$th-order generalization from consecutive derivatives gives Eq.\ (\ref{Resp_0}).

\subsection{Basis-set independent self-consistent free energy response}

To derive the basis-set independent response of the free energy in Eq.\ (\ref{Resp_Omega})
we first calculate the first order derivative of 
\begin{equation}
\Omega_{\rm SCF}[D] = 2Tr[hD] + Tr[DG(D)] - 2 T_e {\cal S}[D^\perp ],
\end{equation}
with respect to $\lambda$ in Eq.\ (\ref{FirstOrderPRT}), i.\ e.
\begin{equation}\begin{array}{l}
\left.\frac{\partial}{\partial \lambda}\Omega_{\rm SCF}[D]\right\lvert_{\lambda = 0}  \\
= 2Tr[h^{(1)} D + h D^{[1]}] + 2Tr[D^{[1]} G(D)] - 2 T_e k_B \beta Tr [F^\perp {D^{\perp}}^{[1]}] \\
= 2Tr[h^{(1)} D] + 2Tr[(h+G(D))D^{[1]}] -2 Tr [F^\perp {D^{\perp}}^{[1]}] \\
= 2Tr[h^{(1)} D^{[0]}] + 2 Tr[FD^{[1]}] - 2 Tr [Z^T FZ {D^{\perp}}^{[1]}]\\
= 2Tr[h^{(1)} D^{[0]}] + 2 Tr[FD^{[1]}] - 2 Tr [F D^{[1]}] = 2Tr[h^{(1)} D^{[0]}]
\end{array}
\end{equation}
where we have derived the entropy derivative as in Eq.\ (\ref{EntDer}) above, used
the definition of the Fockian, $F = h+G(D)$, applied the congruence transformation
between the orthogonal and non-orthogonal representations, e.\ g. $F^\perp = Z^T FZ$,
and the cyclic permutation under the trace.  Subsequent derivatives, analogous to the 
previous Appendix subsection above, gives Eq.\ (\ref{Resp_Omega}).


\end{document}